\pdfoutput=1
\documentclass[reprint,amsmath,amssymb,aps,pra]{revtex4-1}
\usepackage{amsmath}
\usepackage{graphicx}
\usepackage{dcolumn}
\usepackage{bm}
\begin{document}
\preprint{APS/123-QED}
\title{ Aberration-corrected quantum temporal imaging system }
\author{Yunhui Zhu}
\email{yunhui.zhu@duke.edu} \affiliation{Department of Physics and
the Fitzpatrick Institute for Photonics, Duke University, Durham,
North Carolina, 27708 USA}

\author{Jungsang Kim}
\affiliation{Department of Electrical Engineering and the
Fitzpatrick Institute for Photonics, Duke University, Durham, North
Carolina, 27708 USA}

\author{Daniel J. Gauthier}
\affiliation{Department of Physics and the Fitzpatrick Institute for
Photonics, Duke University, Durham, North Carolina, 27708 USA}

\begin{abstract}
We describe the design of a temporal imaging system that
simultaneously reshapes the temporal profile and converts the
frequency of a photonic wavepacket, while preserving its quantum
state. A field lens, which imparts a temporal quadratic phase
modulation, is used to correct for the residual phase caused by
field curvature in the image, thus enabling temporal imaging for
phase-sensitive quantum applications. We show how this system can be
used for temporal imaging of time-bin entangled photonic wavepackets
and compare the field lens correction technique to systems based on
a temporal telescope and far-field imaging. The field-lens approach
removes the residual phase using four dispersive elements. The group
delay dispersion (GDD) $D$ is constrained by the available bandwidth
$\Delta\nu$ by $D>t/\Delta\nu$, where $t$ is the temporal width of
the waveform associated with the dispersion $D$. This is compared to
the much larger dispersion  $D\gg \pi t^2/8$ required to satisfy the
Fraunhofer condition in the far field approach.

\begin{description}
\item[PACS numbers] 03.65.Wj, 03.65.Ud, 03.67.Hk, 03.67.Mn
\end{description}
\end{abstract}

\maketitle

\section{Introduction}

Quantum communication systems rely on transforming and transmitting
information using quantum systems, such as atoms, trapped ions and
photons
\cite{gisin2002quantum,leibfried2003quantum,gisin2007quantum,kimble2008quantum,raymer2012manipulating}.
It is widely believed that future quantum information system will
consist of more than one type of physics system
\cite{gisin2007quantum,kimble2008quantum,raymer2012manipulating,leibfried2003quantum},
such hybrid quantum connections have been demonstrated between ion
and photon \cite{stute2013quantum} and proposed for atoms and
quantum dots \cite{waks2009protocol}. In hybrid systems, photons are
often used as propagating quantum information carrier, or flying
qubits, to connect different quantum systems
\cite{coupling2011Kim,Kimble1997}. In long-distance quantum
communication between hybrid quantum platforms, the wavelength,
temporal scale and spectral profile of the photonic wave-packet for
the source and target quantum memories and transmission channel are
very different \cite{raymer2012manipulating}. We need an efficient
interface to convert the wavelengths and temporal scales  of the
photonic wave packet to match quantum memories, while preserving the
quantum state. Here, we propose a quantum interface for flying
qubits (photons) using temporal imaging integrated with nonlinear
optical wavelength conversion.

The bridge between different wavelengths has been intensely
investigated in the quantum optics field. Quantum connections are
generated either via broadband entangled photon pair sources
\cite{soller2009pcf}, or via nonlinear frequency conversion of the
photonic wavepacket
\cite{kumar1990quantum,Kumar1992Observe,albota2004efficient,langrock2005highly,vandevender2007quantum,pelc2011long,rakher2010quantum,shahriar2012connecting,mejling2012quantum,mckinstrie2005translation,mckinstrie2005quantum,mcguinness2010quantum}.
Preserving the quantum state is achieved using nonlinear frequency
conversion processes that do not amplify the input state (which adds
noise), such as three-wave mixing (3WM)
\cite{kumar1990quantum,Kumar1992Observe,albota2004efficient,langrock2005highly,vandevender2007quantum,pelc2011long,rakher2010quantum,shahriar2012connecting}
and Bragg-scattering-type four-wave mixing
\cite{clemmen2012towards,mejling2012quantum,mckinstrie2005translation,mckinstrie2005quantum,mcguinness2010quantum}.
In these schemes, the quantum state of the signal beam is
transferred to the idler beam at full conversion without excess
noise \cite{mckinstrie2005quantum}. Additionally, the phase of the
pump beam is impressed onto the generated idler waveform
\cite{mcguinness2010quantum}, enabling engineered phase modulation
of the wavepacket.

At the same time, researchers have been investigating temporal
reshaping of photonic wavepacket, while preserving its quantum
states
\cite{temporalShappingEntangled,kielpinski2011quantum,brecht2011quantum,mckinstrie2012quantum}.
Kielpinski \emph{et al.} propose to use a well designed
frequency-dependent dispersion function and temporal phase
modulation to reconstruct the pulse shape
\cite{kielpinski2011quantum}. McKinstrie \emph{et al.} suggest
reshaping the signal pulse profile using pump pulse that has slight
mismatch in group velocity \cite{mckinstrie2012quantum}. Both
proposed schemes requires tailored dispersion functions that are
highly depended on the details of original pulse shape.

\begin{figure}
  \includegraphics[width=8cm]{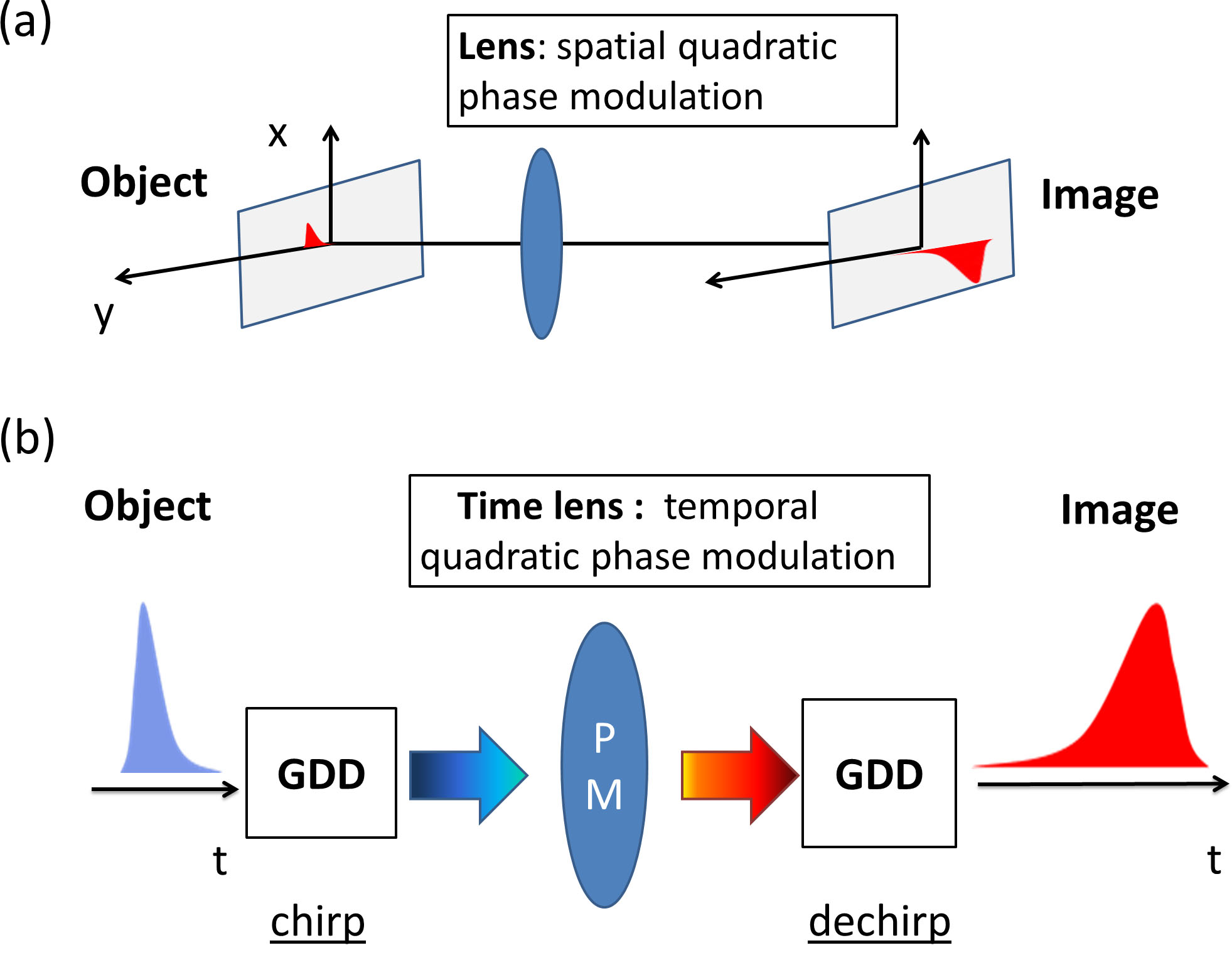}\\
  \caption{Analog of spatial (a) and temporal (b) imaging system.
   GDD: group delay dispersion, PM: phase modulation}\label{0}
\end{figure}

Temporal imaging techniques have been developed by the ultrafast
optics laser community for temporal rescaling of optical pulses
\cite{kolner1989temporal,Tsang2006entanglement,salem2008optical,foster2009ultrafast,foster2008silicon}.
They are the temporal analog of spatial imaging systems. As shown in
Fig. \ref{0}, in a single-lens spatial imaging system, spatial
Fourier components of light waves scattered from the object diffract
into different angular directions. A lens encodes a
quadratically-varying phase to each of these components according to
the direction. The resulting Fourier components then diffract and
recombine to form an image at the image plane. In a temporal imaging
system, temporal Fourier components of light waves are dispersed
into different temporal locations upon propagating in a medium
characterized by a non-zero group velocity dispersion. The dispersed
(or chirped) light wave is modulated by a temporally-varying
quadratic phase, known as a ``time lens.'' Similarly, a temporal
``image'' is formed after a second dispersive medium recombines (or
dechirps) the Fourier components in time.

Here, we combine temporal imaging with a  nonlinear frequency
conversion process that is pumped by a chirped pulse with a
quadratically varying phase profile, which imposes the necessary
time lens phase modulation. In this way, we realize wavelength
conversion and temporal imaging simultaneously.

In a spatial single-lens imaging system, it is well known that a
residual quadratic phase is present at the image even if the
intensity profile is aberration free. Similarly, a residual phase
remains in a single-lens temporal imaging system. In most classical
ultrafast laser applications where only the intensity of the
waveform is detected, residual phase plays no significant role.
However, the phase of the photonic wavepacket is an essential
feature in most quantum applications, such as in conventional
phase-encoded quantum key distribution systems, where the complete
two-dimensional Hilbert space is used for information processing
\cite{gisin2007quantum}. In these applications, it is highly
desirable that this phase be compensated.

In this paper, we present a solution to the residual phase problem
by adding a field lens to the imaging system. We describe the
properties of this imaging system for the case of a time-bin
entangled photonic wavepacket and compare the field-lens technique
with other solutions that are based on temporal telescopes and
far-field imaging. We find that the field-lens approach has better
performance with less dispersion and a simpler setup. The field lens
approach uses only four dispersive elements. The requirement for
group delay dispersion (GDD) $D$ becomes $D>t/\Delta\nu$, where $t$
is the temporal width of the waveform associated with the dispersion
$D$. Compared to the Fraunhofer condition $D\gg \pi t^2/8$ in the
far field approach, the dispersion requirement is dramatically
reduced. As a result, inherent loss in the dispersive elements is
also reduced. This method thus paves the way toward the development
of a highly efficient and flexible flying qubit interface.

\section{Quantum theory for temporal imaging systems}
\begin{figure*}
\centering
  \includegraphics[width=13cm]{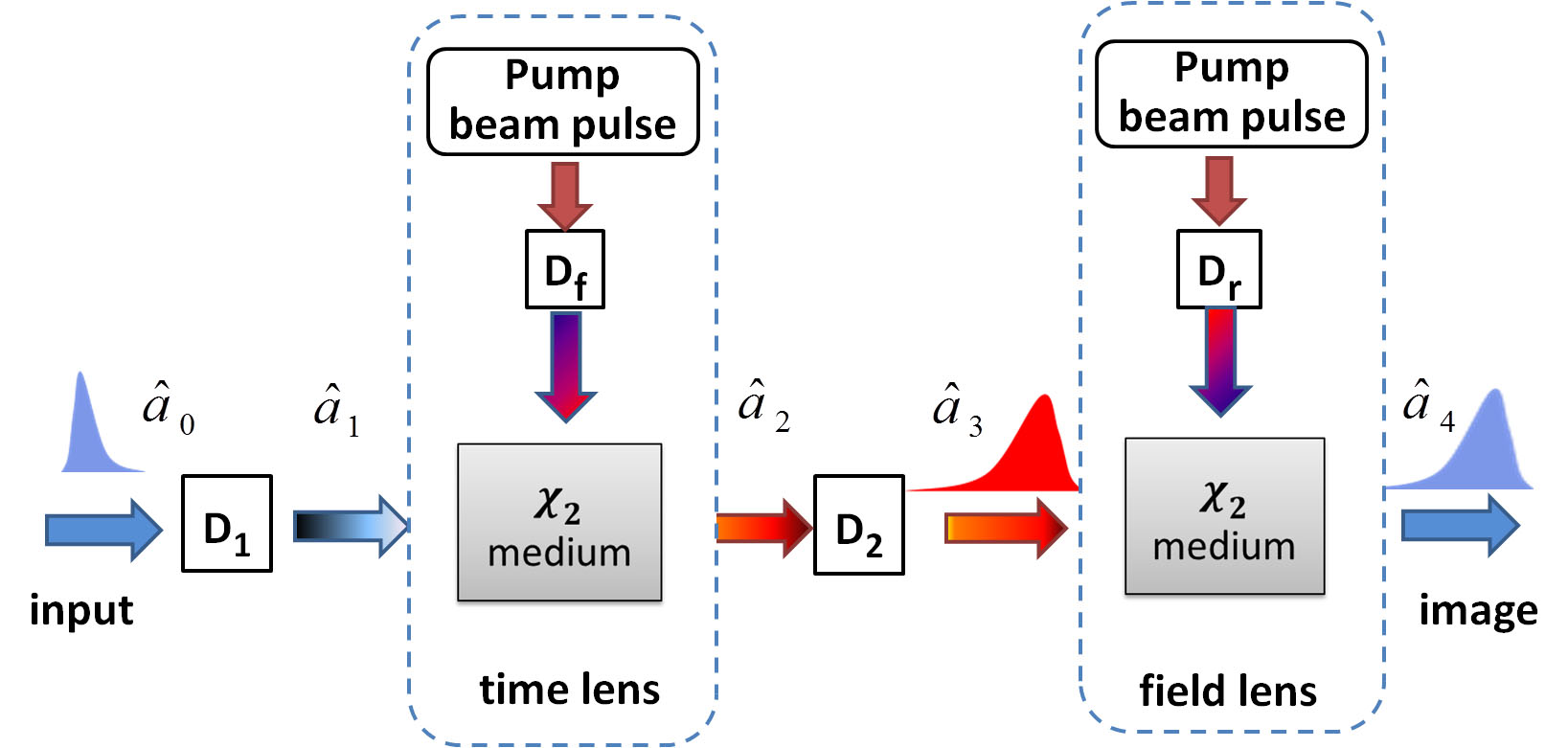}
  \caption{Abberation-corrected flying qubit interface consists of a single-lens imaging system
and a field lens}\label{setup}
\end{figure*}
\subsection{System design overview}
The flying qubit interface consists of a single-lens imaging system
and a field lens placed in the image plane, as shown in Fig. 2. An
input wavepacket (denoted by the annihilation operator $\hat{a}_0$)
first propagates through a dispersive medium $D_1$. The dispersed
wavepacket $\hat{a}_1$ then enters the time lens, which is
constructed using a 3WM process in a crystal with a second-order
nonlinear optical susceptibility $\chi_2$. The beam that pumps the
3WM process (field $E_p$) has a quadratic phase $\phi_p$ obtained
upon propagating through another dispersive material $D_f$, which is
encoded on the generated waveform $\hat{a}_2$ in the 3WM frequency
down-conversion process. The phase-modulated wavepacket $\hat{a}_2$
then propagates through dispersive medium $D_2$ to dechirp (into
$\hat{a}_3$). Finally, a field lens (with quadratic phase $\phi'_p$
on the pump filed $E_p'$ obtained from dispersion $D_r$) frequency
up-converts the wavepacket and removes the residual phase
$\theta_r$. We obtain a chirp free temporal image wavepacket $\hat
a_4$. Here all $D$'s are the group delay dispersion $D=\beta_{2} L$,
where $\beta_{2}$ and $L$ are the group velocity dispersion (GVD)
parameter and length of the dispersive material, respectively.

\subsection{Quantum description of light propagating in a dispersive
material}

To explore the evolution of the annihilation operator $\hat{a}$ for
a photonic wavepacket propagating along the $+z$ direction in a
dispersive material, we expand the operator in the temporal $t$ and
frequency $\omega$ domains as \cite{loudon2000quantum}
\begin{equation}\label{sub}
\hat{a}=\int dt\ \hat a(t)=\int d\omega\ {\hat a}(\omega),
\end{equation}
where $\hat a(t)$ and ${\hat a}(\omega)$ are Fourier transform pairs
and are the temporal and spectral profile annihilation operators of
the mode, respectively. Dispersive propagation is best described in
the frequency domain and is governed by \cite{loudon2000quantum}
\begin{equation}\label{annihilation}
    \frac{\partial \hat{a}(z,\omega)}{\partial z}=\imath\frac{\omega
    n(\omega)}{c}\hat{a}(z,\omega),
\end{equation}
where $c$ is the speed of light in vacuum, and $n$ is the refractive
index of the material. For the case of small dispersion, $\omega
n(\omega)/c$ is expanded around the carrier frequency $\omega_0$ as
\begin{equation}\label{dispersion}
    \omega n(\omega)/c\approx
    \beta_0+\beta_1(\omega-\omega_0)+\frac{\beta_2}{2}(\omega-\omega_0)^2+...,
\end{equation}
where $\beta_i=\partial^i(\omega
n(\omega)/c)/\partial\omega^n|_{\omega=\omega_0}$.

To second-order in the dispersion and ignoring absorption, the
solution to the evolution equation Eq. (\ref{annihilation}) is given
by
\begin{equation}
\hat{a}(z,\omega)=\hat{a}(0,\omega)e^{\imath
(1/2){\beta_2}(\omega-\omega_0)^2z},
\end{equation}
when moving in a reference frame traveling at speed $1/\beta_1$.

We therefore obtain the results that
\begin{equation}\label{0-1}
\hat{a}_1(\omega)=\hat{a}_0(\omega)e^{\imath
(1/2){D_1}(\omega-\omega_0)^2},
\end{equation}
and
\begin{equation}\label{2-3}
\hat{a}_3(\omega)=\hat{a}_2(\omega)e^{\imath
(1/2){D_2}(\omega-\omega_0)^2}.
\end{equation}

\subsection{Quantum theory of the time lens using  the 3WM process}
The time lens is constructed using a 3WM process in a $\chi_2$
crystal. When pumped by a strong (classical) beam
$E_p(t)=A_p(t)e^{\imath\phi_p(t)}$ ($A_p$ and $\phi_p$ being the
amplitude and phase of the pump beam, respectively), the mode
occurrence probability oscillates back and forth between the two
co-propagating signal ($\hat{a}_s$) and idler ($\hat{a}_i$) beams as
a result of sum-frequency and difference-frequency generation (SFG
and DFG). When the phase matching conditions of frequency $\omega$
and wavevector $k$
\begin{eqnarray*}
\omega_i=\omega_s+\omega_p\\
k_i=k_s+k_p
\end{eqnarray*}
are fulfilled, the Hamiltonian for the process is expressed as
\cite{boyd2003nonlinear}
\begin{equation}
H=\int dt\ {\gamma E_p(t){a}_s^\dagger(t) \hat{a}_i(t)+ c.c.},
\end{equation}
where the nonlinear coefficient $\gamma$ is proportional to
$\chi_2$. Note the pump is assumed to not be depleted and thus $E_p$
remains unchanged throughout the process.

Solving the evolution equations of the wavepacket operators, namely
\addtocounter{equation}{1}
\begin{align*}
\partial \hat{a}_s/\partial z=\imath[\hat{a}_s,H], \\
\partial \hat{a}_i/\partial z=\imath[\hat{a}_i,H],
\end{align*}
we find that the mode occurrence probabilities oscillate with the
pump amplitude. Particularly, when the $A_p$ reaches the critical
value so that $ \gamma A_p L_c = \pi/2$, where $L_c$ is the length
of the crystal, the conversion efficiency becomes 100\% and the
optical field switches between two frequency modes, as expressed by
\begin{align*}
\hat{a}_i(z,t)=\imath\hat{a}_s(0,t)e^{\imath\phi_p(t)}\\
\hat{a}_s(z,t)=\imath\hat{a}_i(0,t)e^{-\imath\phi_p(t)}
\end{align*}
By using such a non-amplifying process, the quantum state of the
input signal waveform is transformed to the idler beam (or vice
versa) and the phase (or conjugated phase) from the pump pulse is
imposed onto the output beam as well
\cite{mckinstrie2005quantum,mckinstrie2005translation}.

The result is applied to the time lens sections in the imaging
systems. For the down-conversion time lens
\begin{equation}\label{1-2}
\hat{a}_2(t)=\imath e^{-\imath \phi_p(t)}\hat{a}_1(0,t),
\end{equation}
and for the up-conversion field lens
\begin{equation}\label{3-4}
\hat{a}_4(t)=\imath e^{\imath \phi_p'(t)}\hat{a}_3(0,t).
\end{equation}

\subsection{A single-lens temporal imaging system with residual phase}

We now consider the single-lens temporal imaging system with two
dispersive elements ($D_1$, $D_2$) and a time lens (characterized by
$D_f$). The pump pulse $E_p(t)=A(t)e^{\phi_p(t)}$ has a quadratic
phase $\phi_p(t)=t^2/2D_f$, generated via propagating a short pulse
through a dispersive element with total dispersion $D_f$
\cite{boyd2003nonlinear}.

Combining Eq. (\ref{0-1}-\ref{2-3}) and  Eq. (\ref{1-2}), the output
wavepacket $\hat{a}_3(\omega)$ at the image plane is expressed in
terms of input wavepacket $\hat{a}_0(\omega)$ by
\begin{eqnarray}\label{integration}
  \nonumber  \hat{a}_3(\omega)=&e&^{\imath[D_2(\omega-\omega_0)^2/2]}\int dt e^{-\imath \omega t}e^{-\imath t^2/2D_f}\\
  &\int& d\omega'e^{\imath\omega't}\hat
  a_0(\omega')e^{\imath[D_1(\omega'-\omega_0)^2/2]},
\end{eqnarray}
Carrying out the integration over $t$, we obtain
\begin{eqnarray}
\nonumber
\hat{a}_3(\omega)=&e&^{\imath[D_2(\omega-\omega_0)^2/2]}\\&\int&
   d\omega'\ e^{\imath[-D_f(\omega-\omega')^2/2+D_1\omega'^2/2]} \hat a_0(\omega').
\end{eqnarray}
When the imaging conditions
\begin{eqnarray}\label{imagingcondition}
\nonumber  1/D_1+1/D_2&=&1/D_f\\
 -D_2/D_1&=&M
\end{eqnarray}
are fulfilled, taking a Fourier transform of $\hat{a}_3(\omega)$ and
carrying out the integration over $\omega$, result in the simplified
expression
\begin{equation}\label{a3}
\hat{a}_3(t)=\frac{\imath}{\sqrt{M}}\hat
a_0(0,\frac{t}{M})e^{\imath\theta_r(t)},
\end{equation}
where the output temporal profile is magnified by a factor $M$ and a
quadratic phase $\theta_r=t^2/(2 M D_f)$ is left in the waveform.

\section{Residual phase correction schemes and comparison}

\begin{figure}
\centering
  \includegraphics[width=8cm]{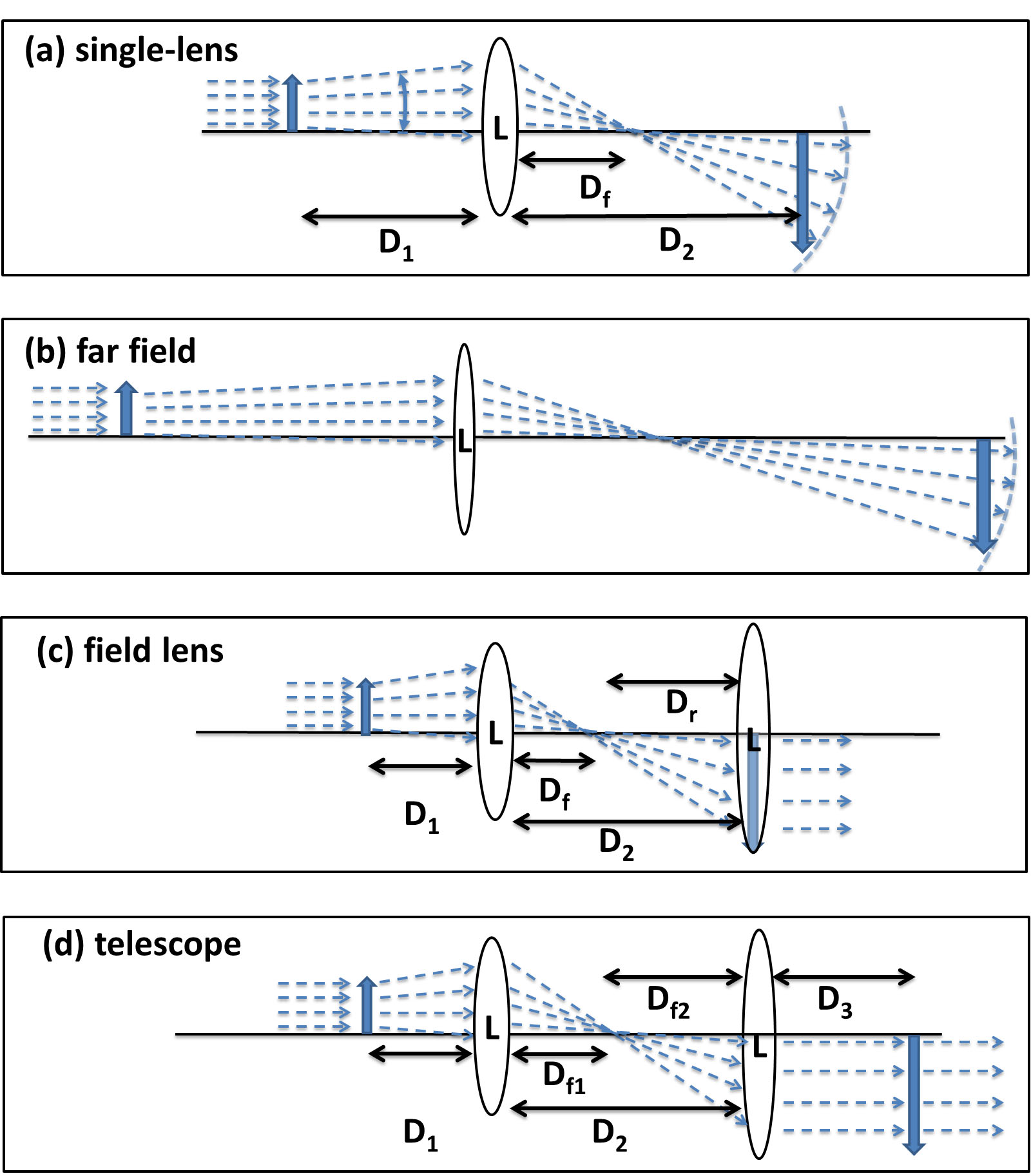}\\
  \caption{Residual phase in a spatial imaging system and three configurations to correct for it.
   Here, $D$ represents diffraction, which is proportional to the distance between each
  elements. Figure (a) shows the
  original single-lens system with image-plane curvature, (b) far-field imaging with reduced
   Petzval field curvature,
  (c) an imaging system with a field lens in the image plane and
  (d) a telescope imaging systems with no field curvature.}\label{compare}
\end{figure}

The quadratic residual phase $\theta_r$ results from the temporally
curved wave front at the image plane, analogous to the spatial image
aberration known as Petzval field curvature (shown in Fig.
\ref{compare}(a)). Petzval field curvature cannot be corrected in a
single-lens imaging system, while other types of abberation such as
spherical aberration can be corrected by a well-designed lens
\cite{born}. Similarly, we have quadratic residual phase in a
single-lens temporal imaging system. (Note that astigmatism and coma
do not appear in a temporal imaging system due to its 1-D nature.)
Since phase does not affect the intensity profile, it is not often
considered in classical applications. Residual-phase free temporal
imaging is discussed by using a telescope in Ref.
\cite{foster2009ultrafast}. In quantum information processing, it is
important to faithfully preserve the phase profile of photonic
wavepacket.

\subsection{Three configurations to solve the residual phase problem}
One method is to reduce the residual phase using larger dispersions,
equivalent to ``far-field imaging'' in spatial imaging systems. In
far-field imaging, the variation of $\theta_r$ is reduced across the
output waveform duration time as a result of the reduced curvature
of the wavefront at the image plane, shown for the analogous spatial
imaging system in Fig. \ref{compare}(b).

A possible method to fully correct for the residual phase is to
include a second lens, known as a field lens, in the image plane as
shown in Fig. 2 and Fig. \ref{compare}(c). Note that we can set the
field lens at the temporal image plane while still spatially
separating the imaged waveform from the lens via non-dispersive
propagation. If the pump pulse of the field-lens  is dispersed by an
amount of $D_r=MD_f$, a phase modulation $\phi_{p}'(t)=-\theta_r(t)$
will be generated and imposed on the wavepacket. The resulting
output image wavepacket is given by Eq. (\ref{a3}) and Eq.
(\ref{3-4}) as
\begin{equation}\label{a4}
\hat{a}_4(t)= e^{\imath
\phi_{p}'(t)}\hat{a}_3(t)=\frac{1}{\sqrt{M}}\hat a_0(0,\frac{t}{M}),
\end{equation}
where a constant $\pi$ phase is ignored. We see that the residual
phase is eliminated, the quantum state of the input field is
transferred to the output field and the temporal profile is extended
by the factor $M$. The wavelength of the single photon will be
properly converted by choosing the appropriate crystal and pump-beam
carrier wavelength.

A third approach is to use the temporal telescope system
\cite{foster2009ultrafast}, as shown in Fig. \ref{compare}(d). This
configuration consists of two time lenses (with focal dispersions
$D_{f1}$ and $D_{f2}$ respectively) and three dispersive elements
($D_1$, $D_2$ and $D_3$). With similar derivation, we find that an
image with magnification $M$ is formed when
\begin{eqnarray}\label{telescope}
\nonumber    D_1=-D_{f1},\\
        D_3=-D_{f2}=-MD_1,\\
\nonumber    D_2=D_1+D_3.
\end{eqnarray}
The output wavepacket  is given by
\begin{equation}\label{a5}
\hat{a}_{4}(t)= \frac{1}{\sqrt{M}}\hat a_0(0,\frac{t}{M}),
\end{equation}
where the residual phase is eliminated.

\subsection{Comparison of three configurations with a single Gaussian pulse}
We now analyze the imaging of a single Gaussian pulse using the
three configurations and compare numerically the required dispersion
and bandwidth. Consider an input single-photon wavepacket with a
Gaussian profile
\begin{equation}
\hat{a}_{0}(t)= \hat\alpha(t) e^{-2ln{2}(t/t_i)^2},
\end{equation}
where $\hat\alpha(t)$ is the annihilation operator of the quantum
mode, $t_i$ is the full width at half maximum (FWHM) of the input
temporal profile. In the single-lens imaging system, according to
Eq. (\ref{a3}), the output photonic wavepacket is
\begin{equation}
\hat{a}_3(t)=\hat\alpha(t) e^{-2ln{2}[t/(M t_i)]^2}e^{\imath t^2/(2
M D_f)}.
\end{equation}
The width of the pulse is expanded to $t_o=Mt_i$. The residual phase
$\theta_r= t^2/(2 M D_f)$ varies by an amount of
\begin{equation}
\Delta\Theta=M t_i^2/(8 D_f)
\end{equation}
over the temporal duration of the output waveform $t_o$. The
quadratic phase can be neglected when\cite{leaird2011dispersion}
\begin{equation}
|\Delta\Theta|\ll\pi.
\end{equation}

Achieving this far-field criterion or Fraunhofer condition requires
that $|D_f|\gg\pi M t_i^2/8$. According to Eq.
(\ref{imagingcondition})  we require that $|D_1|\gg\pi (M+1)
t_i^2/8$ and $|D_2|\gg\pi t_o^2/8$. These dispersion values quickly
become large with increasing output temporal width $t_o$. For
example, if a 100-ps pulse is desired at the output, we require
$|D_2|\gg3,900$ ps$^2$. Note that $3,900$ ps$^2$ corresponds
approximately to the total dispersion of a 200-km SMF-28 optical
fiber at 1550 nm. The required dispersion needs to be much larger,
which is difficult to realize despite various efforts to make
large-dispersion devices for narrow band optical pulses. Popular
approaches include virtually imaged phased arrays (VIPA)
\cite{lee2006optical}, multimode dispersive fibers
\cite{diebold2011giant}, chirped volume holographic gratings
\cite{loiseaux1996characterization} and chirped fiber grating
\cite{littler2005effect}. Nevertheless, it is challenging to
obtaining a total dispersion exceeding $1,000$ ps$^2$, which is
often accompanied by non-ideal characteristics, such as high loss,
higher-order dispersion and group-delay ripple
\cite{shverdin2010chirped,littler2005effect}. For example, a 200-km
SMF-28 fiber at 1550 nm has a transmission of only $10^{-4}$. Such
huge loss will have serious consequences for successful quantum
state transfer. These challenges limit this aberration correction
method to applications requiring pulse in the ps range or shorter.

On the other hand, according to Eq. (\ref{a4}) and Eq. (\ref{a5}),
the output wavepacket $\hat a_o$ in both the field lens and the
telescope configurations are given by
\begin{eqnarray}\label{a4anda5}
\hat{a}_4(t)= \hat\alpha(t) e^{-2ln{2}[t/(M t_i)]^2}.
\end{eqnarray}
eliminating the residual phase independent of the scale of the
dispersions.

As a result, arbitrarily small dispersions can be used until
bandwidth broadening induced by strong (heavily chirped) time lenses
hits the bandwidth limit. In spatial imaging systems, we can move
all components closer (less diffraction) and maintain good imaging
with shorter focal-length lenses. Similarly, systems built with
smaller dispersions require larger quadratic phase modulation.
However, strong phase modulation will expand the spectral bandwidth
of the optical pulses, which may eventually exceed the available
bandwidth of the pump source and/or bandwidth of the 3WM process.
The practical bandwidth $\Delta\nu$, therefore, determines the limit
of the dispersions in these two temporal imaging configurations,
which is much reduced compared to the far-field approach. The
spectral bandwidth of the chirped pump pulse is estimated by taking
the Fourier transform of the pump waveform. Assuming a Gaussian pump
pulse with temporal width $t_i$ and a quadratic phase $\phi_p(t)$
described by
\begin{equation}\label{pump pulse}
    E_p(t)=e^{-2\ln(2)t^2/t_i^2}e^{\imath t^2/2D_f},
\end{equation}
the spectral bandwidth (FWHM) of this pulse is $\Delta\nu=t_i/D_f$.
(Small dispersion $|D_1|\ll t_i^2$ is assumed, so that the input
signal pulse $\hat{a}_1$ is not significantly broadened and
maintains the temporal width $t_i$.) The lower limit of dispersion
is set by the available bandwidth $|D_f|> t_i/\Delta\nu$. Limits for
the other dispersions are obtained via Eq. (\ref{imagingcondition})
and summarized in Table 1.

\begin{table*}[hbt]
\caption{List of dispersion and bandwidth requirements for the
imaging of a single Gaussian pulse (temporal width FWHM=$t_i$) using
three configurations.}
\begin{ruledtabular}
\begin{tabular}{clc|clc|clc}
 \multicolumn{3}{c|}{far field} & \multicolumn{3}{c|}{ telescope}& \multicolumn{3}{c}{field
 lens}\\\hline
   location&dispersion &bandwidth&location &dispersion &bandwidth &location &dispersion &bandwidth\\
\hline
$D_1$ &$ \gg\pi(M+1)t_i^2/8$&$>4\ln(2)/t_i$   &   $D_1$ & $>t_i/\Delta\nu$ & $>4\ln(2)/t_i$ &  $D_1$ & $>\frac{(M+1)t_i}{M\Delta\nu}$ &$>4\ln(2)/t_i$ \\

$D_f$ &$\gg\pi M t_i^2/8$ & $>4\ln(2)/t_i$    &   $D_{f1}$ & $>t_i/\Delta\nu$ & $\Delta\nu$ &   $D_f$ &$>t_i/\Delta\nu$ &$\Delta\nu$ \\

$D_2$ & $\gg\pi M^2 t_i^2/8$ & $>4\ln(2)/Mt_i$  &   $D_2$ & $>\frac{(M+1)t_i}{\Delta\nu}$ & $\Delta\nu$ &   $D_2$ &$>(M+1)t_i/\Delta\nu$&$\Delta\nu$ \\

 & &  &   $D_{f2}$ & $>Mt_i/\Delta\nu$ & $\Delta\nu$    &   $D_r$ & $>Mt_i/\Delta\nu$ & $\Delta\nu$ \\

&&  &   $D_{3}$ & $>Mt_i/\Delta\nu$ & $>4\ln(2)/Mt_i$  &   &&\\
\end{tabular}
\end{ruledtabular}
 \end{table*}

As an example, consider magnifying a 5-ps input waveform at 710 nm
to 100 ps. Pump pulses of bandwidth $\Delta\nu=1\times10^{12}$ rad/s
(roughly twice the spectral width of the input pulse) at 1550 nm are
used as temporal lenses. The input signal is first converted to 1310
nm, and after $D_2$, converted back to 710 nm via the field lens. In
this configuration, the required dispersions are
\begin{itemize}
  \item $D_1$@710 nm: 5.25 ps$^2$,
  \item $D_2$@1310 nm: 105 ps$^2$,
  \item $D_f$@1550 nm: 5 ps$^2$
  \item $D_r$@1550 nm: -100 ps$^2$.
\end{itemize}
The largest dispersion is 105 ps$^2$, well within reach for typical
dispersion devices. These parameters can be obtained using the
following off-the-shelf fiber-based dispersive components:
\begin{itemize}
\item $D_1$, $ 73$ m of SM600 fiber,
\item $D_2$, $ 6.2$ km of LEAF fiber,
\item $D_f$, $ 0.13$ km of VascadeS1000 fiber,
\item $D_r$, $ 5.5$ km of SMF28 fiber.
\end{itemize}
A input wavepacket will go through dispersion material $D_1$
(loss=0.7 dB) and $D_2$ (loss=2.1 dB), with total loss of 2.8 dB. We
see that the system now has much less loss, which can be further
reduced using special low-loss dispersion compensation fiber for
1310 nm and 710 nm.

A similar procedure is used to analyze the telescope configuration.
The results of the dispersion and bandwidth bounds are listed in
Table 1. We find that the telescope configuration uses similar
dispersions as the field-lens configuration. In both cases, the
largest dispersion is $|D_2|>(M+1)t_i/\Delta\nu$, substantially
lower compared to the far-field criterion ($|D_2|\gg\pi t_o^2/8$).
The telescope system requires one additional large dispersion
element $D_3$ compared to the field lens approach which achieves
complete residual phase correction with fewer components and less
loss.

\section{Application: quantum temporal imaging of a time-bin entangled state}

\begin{figure}
  \centering
  \includegraphics[width=8.5cm]{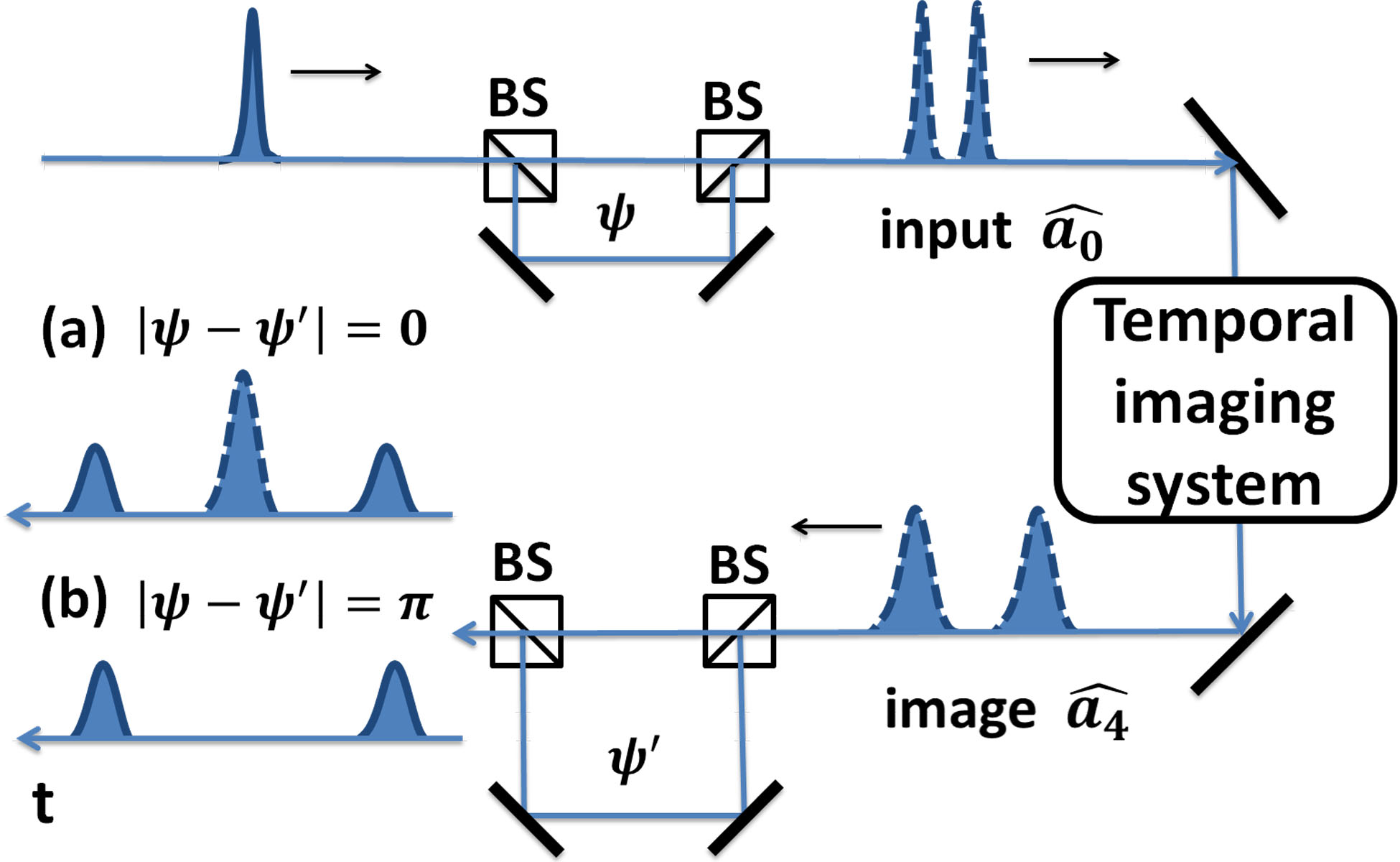}\\
\caption{Setup for preparation, temporal imaging and detection of a
time-bin entangled photonic wavepacket. Constructive interference
(a) and destructive interference (b) in the central peak denote the
time-bin entangled state. BS=50/50 beam splitter.}\label{time-bin}
\end{figure}

As an example application of a quantum temporal imaging system, we
consider a time-bin entangled coherent photon wavepacket, prepared
by splitting a coherent faint laser pulse using a Franson
interferometer (shown in Fig. \ref{time-bin}). The input wavepacket
is a coherent state with a double-Gaussian profile given by
\begin{equation}\label{object}
   \hat{a}_0(t)=\frac{1}{2} A_+(t)\hat{\alpha}(t)
   +\frac{1}{2}e^{\imath \psi}A_-(t)\hat{\alpha}(t)
\end{equation}
where
\begin{equation}
\begin{split}
A_\pm(t)=e^{-2ln{2}((t/\tau\pm d)^2},\\
 \hat{\alpha}(t)=\exp(\hat{a}^\dag-\hat{a}),
\end{split}
\end{equation}
is the coherent state operator, $\tau$ is the width of the Gaussian
profile (FWHM), $\Delta t=2d \tau$ is the propagation delay in the
interferometer, and $\psi$ is the phase difference between the
entangled time bins. The total temporal width of the pattern can be
defined as $t_i=\Delta t+\tau$.

The image waveform of the time-bin entangled photonic wavepacket is
given by Eq. (\ref{a4})
\begin{equation}\label{output}
   \hat{a}_4=\frac{1}{2\sqrt{M}} A_+(t/M)\hat{\alpha}(t)
   +\frac{1}{2\sqrt{M}}e^{\imath\psi} A_-(t/M)\hat{\alpha}(t).
\end{equation}
As shown in Fig. \ref{time-bin}, we split and recombine the output
image wavepacket through another unbalanced interferometer, where
the time difference and phase between the two paths are adjusted to
$M \Delta t$ and $\psi'$. The output temporal waveform is expected
to be a three-peak profile with interference in the central peak.
Constructive interference happens when $|\psi-\psi'|=0$, while
destructive interference happens when $|\psi-\psi'|=\pi$. Since
interference pattern crucially depends on the phase, a complete true
imaging of the phase information encoded in the original time-bin
qubit requires that the residual phase is small throughout the image
temporal profile.
\begin{figure}
  \centering
  \includegraphics[width=8cm]{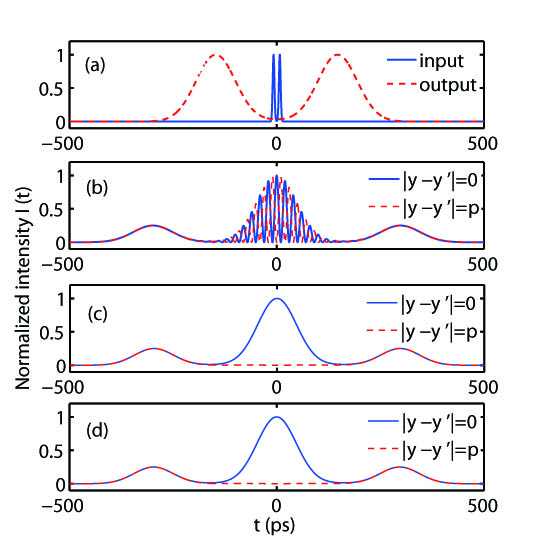}\\
\caption{Interference pattern of a temporal imaged time-bin
entangled photon wavepacket. (a) Input waveform and perfect image
waveform. (b)-(d) Interference pattern after the second
interferometer, simulated for the single-lens imaging system (b),
telescope system (c) and field lens system (d). Blue solid line
shows the constructive interference pattern when $|\psi-\psi'|=0$,
red dashed line shows the destructive interference pattern when
$|\psi-\psi'|=\pi$.  The visibility $v$ is calculated for the
central peak. We obtain $v=0.984$ for field lens and $v=0.986$ for
the telescope system, while in the single lens system, fast varying
residual phase washes out the visibility. }\label{interference}
\end{figure}
\begin{figure}
  \centering
  \includegraphics[width=9cm]{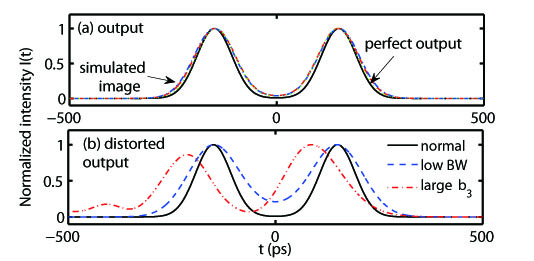}\\
\caption{ Intensity profile aberrations. (a) Simulated image
waveforms using the single-lens imaging system (blue dot-dash),
telescope system (green dot) and field lens system (red dash), which
are almost identical to each other and closely match the expected
perfect output waveform (black solid). (b) Output waveforms in two
severe-distorted situations, insufficient bandwidth $\tau_p=\tau$
(red dash) and large third-order dispersion $\beta_3/\beta_2=1$
ps$\sim\tau$ (blue dot-dash), compared to the expected perfect
output waveform (black solid).}\label{distortion}
\end{figure}

We numerically simulate the evolution of the waveform using Eq.
(\ref{integration}). In the simulation, we set $\tau=5$ ps, $t_i=20$
ps, pump pulse initial width $\tau_p=0.5\tau$ (pump spectrum is
twice as large as the input signal spectrum), and $M=20$, We
simulate the image waveform \emph{intensity} profile
$I(t)=\langle\hat a(t)a(t)^\dagger\rangle$ and the interference
pattern for the single-lens system, telescope system, and the field
lens system. The largest dispersion in each system is restricted to
$Mt_i^2/8=1,000$ ps$^2$. The output intensity profiles are shown in
Fig. \ref{interference}(a) and Fig. \ref{distortion}(a). As shown in
the figure, good image waveform profiles  are formed regardless of
the configuration. Residual phase aberration does not affect the
intensity profile of the image waveform, as expected.

The interference results for the time-bin qubit are shown in Fig.
\ref{interference}(b-d). We see that the interference patterns are
quite different due to the residual phase. In the single-lens
system, the residual phase is large over the temporal profile and
washes out the the interference in the central peak (Fig.
\ref{interference}(b), visibility $v=0$). The field lens (Fig.
\ref{interference}(c)) and telescope configuration (Fig.
\ref{interference}(d)) both have successfully removed the residual
phase, resulting in a high interference visibility.

We also consider two non-ideal factors that causes possible
distortions to the waveform in the simulation: pump-intensity
variation and third-order dispersion. The Gaussian profile of the
dispersed pump beam pulse has intensity variation, which  reduces
the conversion efficiency in the side wings of the input photonic
wavepacket, hence causing distortion of the imaged waveform. A
slight distortion of the waveform towards the center is shown in
Fig. \ref{distortion}(a). Such distortion becomes serious when the
pump pulse does not have sufficient bandwidth, as shown in Fig.
\ref{distortion}(b) for $\tau_p=\tau$. The distortion is reduced by
increasing the bandwidth of the pump pulse, which flattens the
intensity variation.

The dotted-dash line in Fig. \ref{distortion}(b) shows how
higher-order dispersion distorts the quadratic nature of the phase
modulation and causes aberration in the temporal imaging system,
similar to spherical aberration in spatial imaging system. It
becomes serious when $\beta_3/\beta_2>\tau$, and introduces
asymmetric distortion to the waveform. In the simulation, we use
$\beta_3/\beta_2=1$ ps$\sim\tau$. This aberration is not apparent
when we use a value for third-order dispersion
$\beta_3/\beta_2\sim0.1$ ps$<<\tau$ that is appropriate for a
single-mode fiber SMF-28 at 1550 nm.

\section{Conclusion}
We demonstrate a quantum temporal imaging system that allows us to
simultaneously match the wavelengths of two quantum memories and
match their characteristic time scales, enabling exchange of quantum
information between different quantum platforms such as quantum dots
and ions. A field lens in the image plane eliminates the residual
phase in the temporal imaging system. When applied to a time-bin
entangled photonic wavepacket, the image waveform has good
interference visibility, which demonstrates that the field lens
configuration is a good candidate for phase-sensitive quantum
information applications.

We gratefully acknowledge the financial support of the U.S. Army
Research Office MURI award W911NF0910406.

\bibliographystyle{apsrev4-1}
\bibliography{myref}

\end{document}